\documentclass[conference]{IEEEtran}
\usepackage{cite}
\usepackage{graphicx}

\begin{document}
\title{Astrophysical Particle Simulations on Heterogeneous CPU-GPU Systems}
\author{
\authorblockN{Naohito Nakasato\authorrefmark{1}\authorrefmark{2}, 
Go Ogiya\authorrefmark{2}, Yohei Miki\authorrefmark{2} Masao Mori\authorrefmark{2},
and Ken'ichi Nomoto\authorrefmark{3}\\\\
}
\authorblockA{\authorrefmark{1}
Department of Computer Science and Engineering, University of Aizu\\
Aizu-Wakamatsu, Fukushima 965-0815, Japan\\ 
Email: nakasato@u-aizu.ac.jp}
\authorblockA{\authorrefmark{2}
Center for Computational Sciences, University of Tsukuba\\
Tsukuba, Ibaraki, 305-8577, Japan\\
Email: \{ogiya,ymiki,mmori\}@ccs.tsukuba.ac.jp}
\authorblockA{\authorrefmark{3}
Kavli Institute for the Physics and Mathematics of the Universe, University of Tokyo\\
Kashiwa, Chiba, 277-8583, Japan\\
Email: nomoto@astron.s.u-tokyo.ac.jp}
}

\maketitle

\begin{abstract}
A heterogeneous CPU-GPU node is getting popular in HPC clusters.
We need to rethink algorithms and optimization techniques
for such system depending on the relative performance of CPU vs. GPU.
In this paper, we report a performance optimized
particle simulation code "OTOO", that is based on the octree method, for heterogenous systems.
Main applications of OTOO are astrophysical simulations such as 
N-body models and the evolution of a violent merger of stars.
We propose optimal task split between CPU and GPU where GPU is 
only used to compute the calculation of the particle force.
Also, we describe optimization techniques such as control
of the force accuracy, vectorized tree walk, and work partitioning among multiple GPUs.
We used OTOO for modeling a merger of two white dwarf stars
and found that OTOO is powerful and practical to simulate the fate of the process.

\end{abstract}

\IEEEpeerreviewmaketitle

\section{Introduction}
A recent trend to build a cluster system that has heterogeneous CPU-GPU nodes 
makes us to rethink how we develop a high performance code for such system.
From a view point of hardware, there are two important considerations 
to design a heterogeneous CPU-GPU system:
(1) balance between CPUs and GPUs on each node and (2) interconnect architecture.
For instance, the two largest heterogeneous CPU-GPU systems listed in TOP500, 
i.e. Tianjin and TSUBAME2.0, show a different design choice.
Tianjin has 2CPU-1GPU nodes while TSUBAME2.0 has 2CPU-3GPU nodes.
Another recent example is HA-PACS base cluster we used in the present work,
i.e., it has 2CPU-4GPU nodes.
We can compare those clusters with a ratio $R_{\rm GPU}/R_{\rm CPU}$ 
where we define $R_{\rm CPU}$ and $R_{\rm GPU}$ 
the processing speed of double precision operations for CPU and GPU, respectively. 
$R_{\rm GPU}/R_{\rm CPU}$ for the two fastest systems
and HA-PACS base cluster are $3.7, 11.0$ and $8.0$, respectively.
TSUBAME2.0 and HA-PACS base cluster are GPU centric systems while 
Tianjin is a CPU centric system.

The processing speed gained by accelerators is not effectively utilized without
restructuring of algorithms and optimization techniques used in existing parallel codes.
We think the balance between $R_{\rm CPU}$ and $R_{\rm GPU}$ 
highly affects an optimization strategy for codes running on a heterogeneous CPU-GPU system.
Notably, we have two ways to take advantage of a node with multiple GPUs, 
that are adopted by TSUMABE2.0 and HA-PACS base cluster.
One is a flat MPI parallelization that explicitly parallelize a code 
with MPI and assign each MPI process a GPU device.
Alternative way is similar to a hybrid MPI-OpenMP parallelization.
In this way, we launch one process per a node and the process
controls multiple GPU devices simultaneously.
For a given $R_{\rm CPU}$ and $R_{\rm GPU}$,
which approach is optimal is not a trivial question to answer.
In this paper, we report the performance optimization of a octree method, 
that is a powerful algorithm for particle simulations \cite{Barnes_1986}, 
on heterogeneous CPU-GPU systems.
We have reconsidered the all aspects of the octree method
and implemented a new code called OTOO (OcTree On Opencl).

The octree (or $k$-d tree in general) method is a standard algorithm used in many applications.
Here we only introduce notable examples related to astrophysics.
The octree \cite{Barnes_1986} and binary \cite{Jernigan_1989} tree methods
have been heavily used in 
astrophysical simulations to speed-up two important interactions between $N$ particles;
gravity and Lagrangian hydrodynamics, where $N$ is the number of particles.
Despite the $O(N \log N)$ complexity, optimizations
of a tree method by techniques such as vectorization and
parallelization is necessary to accommodate demands for
simulations with larger and larger $N$. 
\cite{Hernquist_1990}, \cite{Makino_1990}, and \cite{Barnes_1990} have reported various
techniques to vectorized the gravity calculation with the octree method. 
\cite{Warren_1992}, \cite{Dubinski_1996}, and \cite{Yahagi_1999}
have reported a parallel tree method for massively parallel processors (MPPs). 
In a pioneering work by \cite{Xu_1995}, they have proposed a unification
of the tree method and the particle-particle particle-mesh method \cite{Hockney_1988}.
This TreePM method is a popular method for large scale cosmological $N$-body simulations
(e.g., \cite{Ishiyama_2009}).
Another computational technique to speed up the tree method
utilizes the GRAPE special-purpose computer \cite{Sugimoto_1990, Makino_1998}. 
Using a combination of vectorization techniques for the tree method, 
the tree method can be executed efficiently on a GRAPE system \cite{Makino_1991}.
This method is very effective for large scale simulations;
\cite{Makino_2004} have reported an parallel implementation of the tree method with GRAPE
and \cite{Yoshikawa_2005} have reported an parallel TreePM with GRAPE.
A similar approach has been used with GPUs in \cite{Hamada_2009, Hamada_2010}.

An effective method to solve evolution of astrophysical plasm
is Smoothed Particle Hydrodynamics (SPH) method proposed by \cite{Gingold_1977, Lucy_1977}.
In \cite{Springel_2001, Wadsley_2004, Springel_2005, Wetzstein_2009},
they have presented own parallel tree code that supports both gravity and SPH interactions.
Although we can not list all relevant results, those parallel gravity+SPH codes 
have been quite frequently used for many problems in astrophysics.
Notably, the gravity+SPH method has been successfully used in modeling 
the fate of a violent merger of two white dwarf stars as in a pioneering work \cite{Benz_1990}
and recent work \cite{Dan_2012, Pakmor_2012, Raskin_2012}.
Since the fate of the process is thought to be closely related to a Type Ia explosion
\cite{Iben_1984,Webbink_1984}
that is a standard candle in the universe and has been used to measure
important cosmological parameters of the universe \cite{Riess_1998, Perlmutter_1999}
(awarded Nobel prize in Physics 2011), 
a large scale modeling of mergers of two white dwarf stars
is currently a very hot topic in astronomy and astrophysics \cite{Dan_2012, Pakmor_2012, Raskin_2012}.
However, there are no gravity+SPH code that takes advantage of a new trend 
of heterogeneous computing.

In this paper, we present our new code OTOO based on the octree
specially optimized for heterogeneous CPU-GPU systems.
By combining our previous work on an implementation of the octree code
on GPU \cite{Nakasato_2011} and our SPH code for astrophysics \cite{Nakasato_2003}
with required modifications, we have developed OTOO code.
We have adopted the OpenCL programming model to implement parallel and vectorized
tree traversal on OpenCL devices, that are usually many-core CPUs or multi-core GPUs.
OTOO code running on a combination of CPUs and recent GPUs
shows competitive performance to a standard tree code running on a MPP.

In the following sections, we will describe details of our approach and optimization techniques.
Section 2 overviews the octree method and relevant data structure.
Our novel optimizations for heterogeneous systems are presented in Section 3.
We show our performance tests on various configurations in Section 4. 
In Section 5, we present the modeling of a merger of two white dwarf stars
followed by discussion and summary in Section 6.


\section{Octree Method}
The octree method \cite{Barnes_1986} is a special case of the general $k$-d tree algorithm. 
This method is optimized to efficiently calculate the mutual interactions between particles, 
and reduces the computational complexity of the force calculation from
$O(N^2)$ for the brute force method to $O(N \log N)$ where $N$ is the number of particles. 
It approximates the force from a group of distant particles using their multipole expansions.
Note that there is a trade-off between the approximation error 
and the way in which we compute the force through multipole expansions. 
A tree structure that contains all particles is used to judge this
trade-off efficiently.
The octree method is executed in two
steps: (1) a tree construction and (2) a force calculation. 

In the tree construction, we divide a cube that encloses all particles into eight equal sub-cells. 
The cell is the root of a tree that we construct; it is called the root cell. 
Then, each sub-cell is recursively subdivided in the same say until 
each cell contains particles less than a specified number of particles.
As the result of this procedure, we obtain a tree of cells and particles.
Note that the number of maximum particles in a cell, hereafter $n_{\rm crit}$,
is an important parameter that determines the depth (or the height) of the tree structure.
Larger $n_{\rm crit}$, we have a shallower tree and it takes shorter time to construct it.
The original paper \cite{Barnes_1986} has adopted $n_{\rm crit} = 1$.
As the result, the constructed tree structure is a complete octree.
With $n_{\rm crit} > 1$, the constructed tree is no longer a octree since 
cells at the bottom of the tree can contain more than 8 particles.

After the tree is constructed, we traverse it to judge whether
we should replace a distant cell that contains a group of 
particles with the multipole expansions of those particles. 
The way how we judge the trade-off is called a multipole acceptance conditions (MAC).
If we do not replace/accept the cell with its multipole expansions,
we then further traverse sub-cells of the distant cell. 
If we do replace/accept the cell, we calculate a particle--cell interaction. 
In this case, we normally compute the gravity force using the multipole expansions
of the order $p = 1$ (monopole) to $p = 3$ (quadrupole).
In the present work, we only consider the monopole component so that
the particle--cell interaction is a simple interaction between 
the particle and a virtual particle at the center of the mass of the cell.
When we encounter a particle during the tree traversal,
we immediately calculate a particle--particle interaction. 
With an appropriate MAC selected, the computational complexity
is reduced to $O(N \log N)$ from $O(N^2)$ when we do not accept cells at all.

To implement a tree code, we need to specify a data structure
and a MAC that controls the approximation error in the force.

\subsection{Tree Representation}
In \cite{Castro_2008}, they have summarized data structures to represent the tree;
(1) exhaustive tree representation, (2) hashed octree and (3) bother/child representation.
The first representation used in \cite{Barnes_1986} is
that each cell has eight pointers to own child cells.
The hashed octree \cite{Warren_1993} assigns a key to each cell
and uses a hash table to efficiently access the cell data.
The relation between parent and child cells is encoded in the Morton key \cite{Morton_1966}.
The brother/child representation is essentially a linked-list; 
a cell has two pointers to the next brother cell and to the first child cell.
This representation has been used by \cite{Makino_1990a} with a slight modification.
The brother/child representation is flexible data structure 
since it can represent any complex tree structure not only a mere octree.
Furthermore, the brother/child representation allow us 
to traverse a tree {\it iteratively} \cite{Makino_1990a}
while we normally traverse a tree {\it recursively} 
with the exhaustive and hashed tree representation.
This nature of the brother/child representation is critical 
to implement tree traversal on GPU \cite{Nakasato_2011}.
Note with a help of stack it is possible to iteratively
traverse a tree with the exhaustive and hashed tree representation
on GPU as proposed by \cite{Bedorf_2012}.

\begin{figure}
\centering
\includegraphics[width=8.5cm]{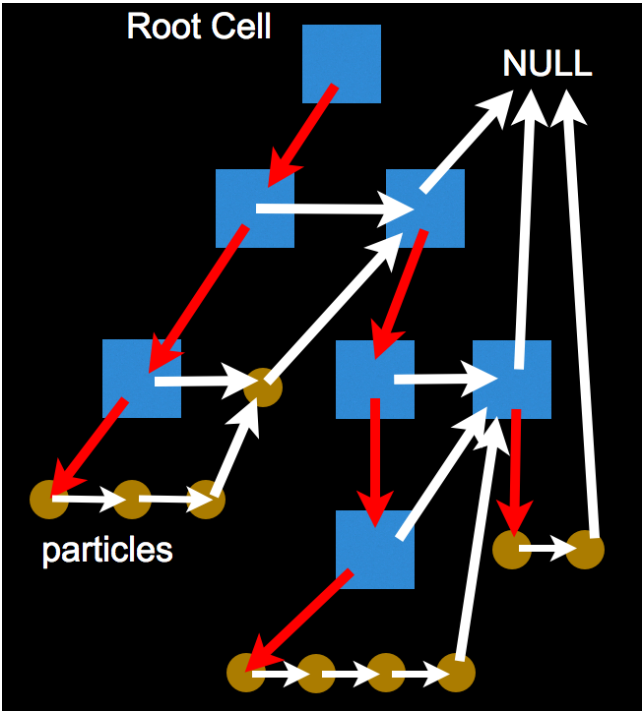}
\caption{
A tree consists of 10 particles and 7 cells.
Red and white arrows show {\it more} and {\it next} pointers, respectively.
}
\label{TREE}
\end{figure}

According to \cite{Makino_1990a}, we call the brother pointer a {\it next} pointer
and the child pointer a {\it more} pointer in the present work.
A reason is that the {\it next} pointer of a cell points the cell's next brother
or the parent's {\it next} pointer, which is normally the parent's next brother,
if the cell is the last brother.
In Figure \ref{TREE}, we present a schematic view of a tree data.
With this modification, we traverse the tree starting from the root cell
as follows;
(1) when we encounter a cell and do not accept the cell, 
we follow the cell's {\it more} pointer and continue.
(2) when we accept the cell, we further traverse the cell pointed by its {\it next} pointer
after computing the particle--cell interaction.
(3) when we encounter a particle, we immediately compute the particle--particle
interaction and then further traverse the cell pointed by its {\it next} pointer.
(4) when we encounter the null pointer, we stop the traversal.

\subsection{MAC Condition}
\begin{figure*}
\centering
\includegraphics[width=13cm]{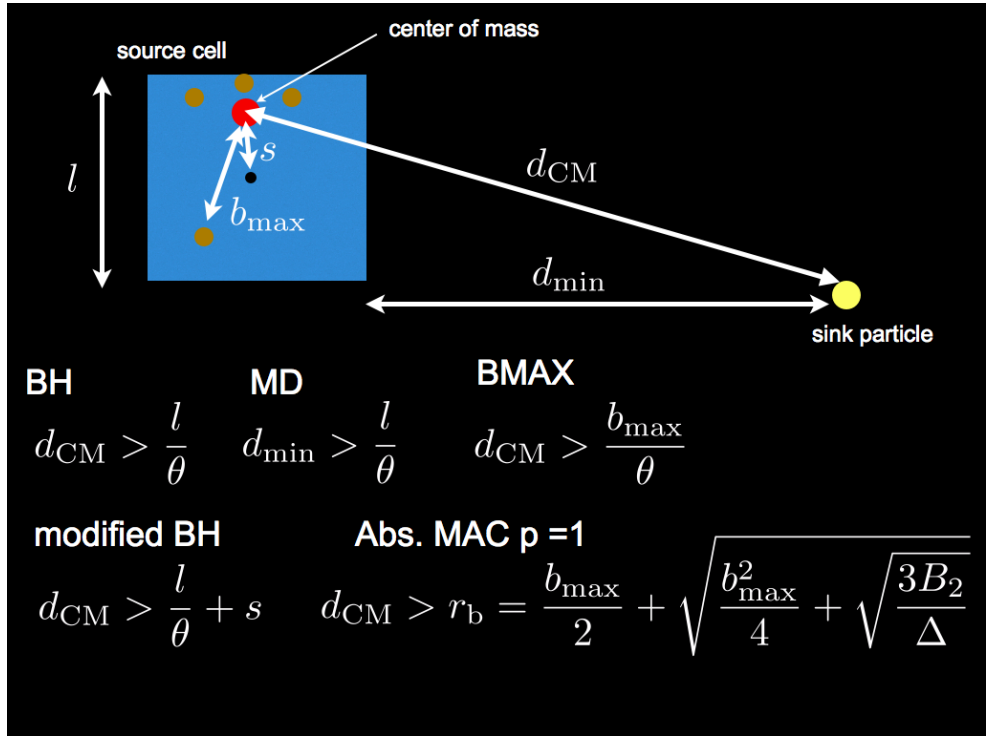}
\caption{
``BH'' is the original MAC proposed by X. 
}
\label{ALLMAC}
\end{figure*}

Beginning from the original simple MAC \cite{Barnes_1986}, 
a several MACs have been proposed by many authors.
In \cite{Salmon_1994}, Salmon and Warren have compared existing MACs 
with the their proposed MACs.
Given a group of particles they have derived the error bounds for multipole expansions
and proposed a several better MACs based on the error bounds.
With the proposed MACs by them, we compare a distance between 
the particle, where we want to compute the force, 
and a cell that contains a group of particles with the bounding radius $r_{\rm b}$ of the cell.
They have derived $r_{\rm b}$ from the analytical error bounds but $r_{\rm b}$ adopted in all other MACs
was defined by only considering geometry and ignoring particle distribution.
Figure \ref{ALLMAC} shows schematic comparison of various MACs.
We use the following notation; 
the size of the cell is $l$, $\vec{r}_{\rm CM}$ is the center of mass, 
$s$ is the distance between $\vec{r}_{\rm CM}$ and the geometrical center of the cell, 
$b_{\rm max}$ is the maximum distance of a particle from $\vec{r}_{\rm CM}$.
We define two distances between the source cell and the sink particle;
$d_{\rm CM}$ is the distance between the sink particle and $\vec{r}_{\rm CM}$
and $d_{\rm min}$ is the minimum distance between the sink particle and the source cell.

In the present work, we use the absolute MAC proposed by same authors in \cite{Warren_1993}.
With this MAC, $r_{\rm b}$ is given as
\begin{equation}
r_{\rm b} = \frac{b_{\rm max}}{2} + \sqrt{\frac{b_{\rm max}^2}{4} + \sqrt{\frac{3 B_2}{\Delta}}}, 
\label{SW_MAC}
\end{equation}
where $\Delta$ is an accuracy parameter with the dimension of acceleration.
In Eq. (\ref{SW_MAC}), $B_2$ is the trace of the quadrupole moment tensor
defined as
\begin{equation}
B_2 = \sum_i^{N_p} m_i | \vec{r}_{\rm CM} - \vec{r}_i |^2,
\label{B2}
\end{equation}
where $n_p$ is the number particles in the cell and $\vec{r}_i$ represents the position of particle $i$.
$B_2$ can be computed by a summation over $n_p$ particles 
after we compute $\vec{r}_{\rm CM}$ during the construction phase.

Suppose we have two particles with mass $m$ separated by $L$ in a cell
so that $b_{\rm max} = L/2$ and $B_2 = 2 m (L/2)^2 =  m L^2/2$.
\begin{equation}
r_{\rm b} > \frac{L}{4} + \sqrt{\frac{L}{16} + \sqrt{\frac{3 m L^2}{2 \Delta}}}
\end{equation}

\begin{figure}
\centering
\includegraphics[width=8.5cm]{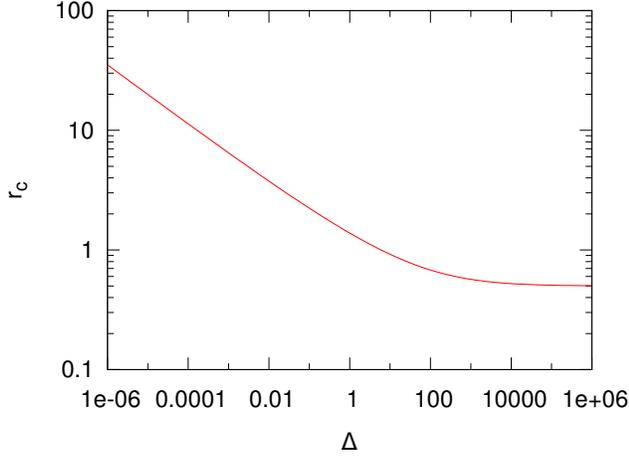}
\caption{
The bounding radius $r_{\rm b}$ for two particles separated 
by 1 as a function of the absolute accuracy parameter $\Delta$.
}
\label{MAC}
\end{figure}

Figure \ref{MAC} shows $r_{\rm b}$ 
as a function of the accuracy parameter $\Delta$ with $m = L = 1$. 
Note that if $\Delta$ is large, $r_{\rm b} \sim b_{\rm max}$ that is the same
as the b$_{\rm max}$ MAC described in \cite{Warren_1993, Salmon_1994}.
On another extreme where we require accurate force with very small $\Delta$, 
the dependence of $r_{\rm b}$ on $\Delta$ is weak $\propto \Delta^{-0.25}$.
Since the number of interactions scales $\propto r_{\rm b}^3$, 
the calculation costs scales $\propto \Delta^{-0.75}$.

\subsection{Bounding Radius for SPH}
To compute the evolution of a merger of two white dwarf stars, 
we extended our octree for SPH method.
Here, we do not describe the details of our SPH implementation 
but we note that in SPH method we need to compute summations over neighbor particles 
such as
\begin{equation}
\langle A(\vec{r}_i) \rangle = \sum A(\vec{r}_j) W(|\vec{r}_j - \vec{r}_i|),
\label{SPH_SUM}
\end{equation}
where $\langle A(\vec{r}_i) \rangle$ is the average value of a physical value $A$
with a given kernel function $W(r)$.
The kernel function $W(r)$ is a bell shape function that is zero outside
of a given neighbor radius $h$.
Namely, all computations relevant in SPH are summation between particles with a given radius $h$.
There are two possible approaches: (1) calculate the neighbor list
and (2) directly evaluate the summations.
With the first method, we can use the tree structure to efficiently select the list of neighbor
particles that are inside a sphere with the radius $h$.
We implemented this procedure called a neighbor search 
using the same tree structure for gravity with a modified bounding radius.
The second method is a strait-forward modification of the octree traversal for gravity.
Although we do not need multipole expansions for SPH interaction, 
the basic principle of acceptance of a cell with a bounding radius is the same
as the gravity.
We have implemented the two method on top of OpenCL.
In the present work, we adopted the second method.
The performance evaluation of the neighbor search algorithm on OpenCL will be presented elsewhere.

We compute the bounding radius of a cell with following simple equation as
\begin{equation}
r_{\rm SPH} = {\rm max}_i^{N_p} |\vec{r}_{\rm CM} - \vec{r}_i| + h_i,
\label{SPH_RAD}
\end{equation}
where $h_i$ is the neighbor radius of $i$-particle. 
${\rm max}_i^{N_p}$ means that we compute the maximum over $N_p$ particles belonging to the cell.
The bounding radius $r_{\rm SPH}$ guaranties that we can select correct neighbor particles.
With this definition, we can manage the gravity and the SPH interactions in an unified manner.
The tree traversal algorithm is the same as that of gravity.
Practically, we directly compute summations like Eq. (\ref{SPH_SUM})
when we encounter a particle along the tree traversal.
There is no particle--cell interaction in SPH interaction.

The use of the tree structure for a neighbor interaction 
is very effective in astrophysical SPH simulations where
particle distribution is normally non-uniform due to self gravity and complex shock waves.
On the other hand, many other work of a SPH implementation on GPU, e.g., \cite{Harada_2007, Krog_2012}, 
were for simulations of quasi-incompressible fluid that exhibits uniform density by nature.
They have adopted a neighbor search scheme using uniform grid with the spacing of $h$.
Note the grid based neighbor search algorithm is only effective
when particles are distributed with roughly uniform density.
Since it is common practice that we make $h$ for each particle variable in astrophysical
SPH simulations, the grid based algorithm is not working effectively.

\section{Optimization of tree method for Heterogeneous CPU-GPU System}
The octree code in the present work is based on the OpenCL kernel
that was presented in our previous paper \cite{Nakasato_2011}.
We choose OpenCL for our implementation since
OpenCL is a standard programming model for both many-core GPU and multi-core CPU. 
The performance of our OpenCL octree code presented in our previous work \cite{Nakasato_2011} 
was not mature in 2010, however, recent development of OpenCL software development kits (SDK)
from a several vendors makes it possible to use them for production runs.
In the present work, we have tested three SDKs from AMD, Intel and NVIDIA.
Only the SDK from AMD supports both CPU and GPU (AMD GPU only) while 
the SDK from Intel supports only CPU and the SDK from NVIDIA supports only own GPU.
Our computing kernels works on all three SDKs without any modifications.
In this section, we describe our optimization strategy and techniques
for OpenCL devices.

\subsection{System Configuration}
First, we describe system configurations used in the present work
since we can not optimize a code without specifying details of the performance of a system.
Table \ref{configuration} lists the configurations we used in the present work.
We have three categories; (1) 2090HAP: a system with multiple GPUs, 
(2) 7970SB, 7970BD and APU: a system with single GPU 
and (3) HAP, OPT, SANDY: a system with only CPU.
2090HAP is a node of HA-PACS base cluster system recently installed at 
Center for Computational Sciences in University of Tsukuba.
It has dual E5-2670 CPU and four M2090 GPU boards.
HA-PACS base cluster consists of 268 nodes interconnected by dual QDR inifiniband.
7970SB and 7970BD are systems with a latest Radeon 7970 board.
APU is a system with the heterogenous CPU (A6-3650) 
that combines CPU cores and GPU cores in a chip.
HAP, OPT and BD are systems in which we use only CPU cores.
For 2090HAP, 7970SB and 7970BD, 
the fourth column indicates the version and the number of lanes of PCI Express interface.
Only 7970SB support the latest Gen.3.0 specification.
The fifth column shows the peak performance in single precision operations in $10^9$ flops.
Note we only put the performance of GPU for the systems with GPU.
Except explicitly stated, we used g++ version 4.4.3 and 4.4.5 (only on 2090HAP and HAP) with
the optimization option ``-O3'' in the present work.

\begin{table*}
\renewcommand{\arraystretch}{1.3}
\caption{System Configurations used in the Present Work}
\label{configuration}
\begin{center}
\begin{tabular}{c|c|c|c|c|c}
\hline
Name   & CPU & GPU & PCIe & SP perf. & SDK \\
\hline
2090HAP & dual Xeon E5-2670 & M2090 x 4 & Gen.2.0 x16 & 5320 & NVIDIA \\
7970SB  & Core i7-3960X     & HD7970    & Gen.3.0 x16 & 3789 & AMD \\ 
7970BD  & FX-8150           & HD7970    & Gen.2.0 x16 & 3789 & AMD \\
APU     & A6-3650           & HD6530D   & -- & 284 & AMD \\
HAP     & dual Xeon E5-2670 & --        & -- & 666 & Intel \\
OPT     & dual Opteron 6168 & --        & -- & 364 & AMD \\
SANDY   & Core i7-3960X     & --        & -- & 316 & Intel \\
BD      & FX-8150           & --        & -- & 230 & AMD \\
\hline
\end{tabular}
\end{center}
\end{table*}

\subsection{Tree Construction}
\label{TC}
In our previous work \cite{Nakasato_2011}, we choose to construct tree structure
on CPU with a single thread.
For $N \sim 0.8$ M case, we have reported the time for the tree construction
took roughly 27\% of the total time
and the fraction of this part was a increasing function of $N$.
In the present work, we still use CPU to construct the tree 
but we switch to a parallel implementation as much as possible.
We selected a modified algorithm used in the hashed octree code by \cite{Warren_1993}.
To be precise, our tree construction algorithm is divided into the following four steps:

\begin{itemize}
\item calculation of the size of the root cell
\item key generation for each particles
\item sort the keys
\item setup the linked-list pointers
\end{itemize}

Except the last step, all steps are possible to implement in parallel.
We skip the description for the first step since it is obvious
as we just compute the size of the box enclosed all particles.

For the second step, our algorithm first computes the Morton key \cite{Morton_1966}
followed by converting it into the Hilbert key.
One way to compute the Morton key from a particle's coordinate $\vec{r}$
is a bit interleaving used in \cite{Warren_1993}.
With a center of a cube cell, it is easy to compute which the octant
is a particle in by interleaving the three bits, each of which represents
the particle is at either side of the center in each axis.
The obtained integer (hereafter octant index) is 0 - 7 and 
is used to construct the octree structure.
A trick is that the Morton key is a concatenation of the octant index at each level of the octree.
Namely, we can readily obtain the octant index from the Morton key
with shift and mask operations.
This nature is preserved even after we convert the Morton key into the Hilbert key.
Alternatively, we use the dilation algorithm presented in \cite{Raman_2007}.
Their algorithm is highly optimized to convert $\vec{r}$ 
into the dilated integer that is identical to the Morton key.
This computation is done in parallel for each particle.

It is possible to construct the octree structure from the Morton key
but a drawback of the Morton key is that it preserves the data locality only approximately.
Namely, we see many discontinues jump in particles sorted in the Morton key.
Due to the fact, we further convert the Morton key into the Hilbert key \cite{Kamata_1991, Bandou_1998}
for better preserving the data locality.
This conversion is also done in parallel.
In both computations, we simply parallelize loops 
for the computations with {\it parallel for} directive in OpenMP.

After computation of the keys, we sort an array of the pair of a key and a particle's id with 
a sort algorithm.
We have tested various sorting algorithms both in serial and parallel.
We found that the sort algorithm in GNU libstdc++ parallel mode \cite{Singler_2007}
that is implemented as a parallel quick sort, is the fastest for our purpose.
This algorithm internally uses OpenMP for parallelization.

Finally, in the last step, we compute the linked-list pointers
directly from the sorted keys in contrast to other works.
The pointers are not an address pointer but an integer index
because we use arrays to store $\vec{r}$, mass $m$ and other properties of a cell or a particle.
Suppose we have $N$ particles, the index from $0$ - $N-1$ is for particles
and the index larger than $N-1$ is reserved for cells.
With the sorted keys, we readily find group of particles 
with the same octant index by slicing the keys. 
The root cell contains all particles by definition.
It has eight sub-cells and we readily find groups of particles in the root cell
with the same octant index.
For the eight groups of particles, 
we create eight new cells and we continue the same procedure recursively for each cell.
We stop to further create a new cell when we have less than $n_{\rm crit}$ particles in a cell.
For those particles in the cell, we link the particles with {\it next} pointers.
After finishing the tree construction, we have a tree that is an octree in shallower cells
and $k$-d tree in bottom cells that contains only particles.
This procedure is difficult to naively parallelize without an atomic operation
for creating a new cell.

\subsection{Update Properties of Cells}
\label{MM}
For each cells, we need to compute a several properties used for the force calculation.
If we use the tree to compute gravity force, 
we need to compute the total mass and the center of mass of a cell.
These properties can be computed by a recursive depth-first tree traversal.
Alternatively, we can compute them by direct summation for each cell.
While the computational complexity of the recursive algorithm 
is better than the direct summation algorithm,
it is difficult to parallelize but the direct summation algorithm is possible to parallelize easily.
By comparing the performance of the two algorithms, 
we found the recursive algorithm is faster in a typical case. 

After computing the center of mass, we then compute
the trace of the quadrupole moment tensor $B_2$
by directly computing Eq. (\ref{B2}) for each cell.
Note that we do not necessarily to compute $B_2$ for massive cells like the root cell.
This is because those massive cells are almost always {\it not accepted} with the MAC we adopted.
We introduce a parameter $m_{\rm limit}$ in our algorithm.
If the mass of a cell is more massive than $m_{\rm limit}$, 
we do not compute $B_2$ and set the bounding radius $r_{\rm b} = 2 S_{\rm root}$
where $S_{\rm root}$ is the size of the root cell.
This apparently large $r_{\rm b}$ makes a massive cell not accepted all the time.
We usually set $m_{\rm limit}$ a few percent of the total mass.
Note that this treatment introduces any additional error in the force calculation.
With $B_2$ for cells that are less massive then $m_{\rm limit}$, 
we compute $r_{\rm b}$ with Eq. (\ref{SW_MAC}).

\subsection{Optimization of Tree Traversal}
In our octree code, we send the following data to OpenCL device;
(1) $\vec{r}$ and $m$ of particles and cells ($4 \times (N + N_{\rm cell})$ words),
(2) $r_{\rm b}$ ($N + N_{\rm cell}$ words), and
(3) {\it next} and {\it more} pointers ($2 \times (N + N_{\rm cell})$ words).
For the force calculation, $r_{\rm b}$ is necessary only for cells
but we use it to store the gravitational softening length for particles.
And it is not necessary to send the {\it more} pointers for particles
since we do not further divide the particle.
But for a simplicity, we set it null value.
In total, we need to send $7 \times (N + N_{\rm cell})$ words.
In the present work, we use single precision variable so that 1 word is 4 bytes.

After sending the required data to a OpenCL device, 
we start the tree traversal on the OpenCL device.
We have implemented the kernel that executes the tree traversal for a particle
in our previous work \cite{Nakasato_2011}.
A vectorized tree code proposed by \cite{Barnes_1990}
executes the tree traversal for multiple particles. 
For a group of particles, they has computed the minimum distance from the group
to cells or a particle and used it for a MAC.
Their definition of the minimum distance is shown in Figure \ref{MAC}
(as indicated by $d_{\rm min}$).
We adopted their {\it vectorized} algorithm with a modification.
This algorithm makes the tree traversal more compute intensive.
It is an effective optimization technique for computation on GPU.
Suppose we have a group of $n_{\rm vec}$ particles for the tree traversal.
During the traversal, we encounter a cell for testing the MAC.
We fist compute $n_{\rm vec}$ distances for each pair
and then select the actual minimum distance 
from $n_{\rm vec}$ distances by binary reduction. 

\begin{figure}
\centering
\includegraphics[width=8.5cm]{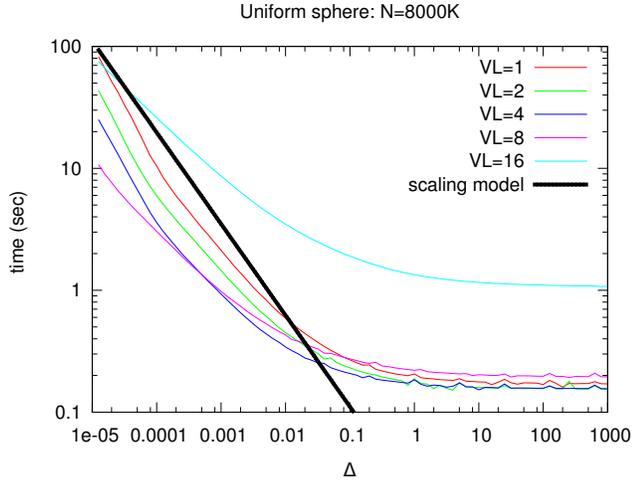}
\caption{
The calculation time with various vector length for the uniform sphere
as a function of the parameter $\Delta$.
}
\label{mac_experiment_time}
\end{figure}

Here, we present the experimental results for the absolute MAC with different $n_{\rm vec}$.
Figure \ref{mac_experiment_time} shows the measured calculation time
on 7970SB system in Table \ref{configuration} as a function of $\Delta$.
In this measurement, we have set up a sphere with uniform density
with radius $r = 1$, total mass $M = 1$, and $N = 8000 \times 1024$; roughly 8 M particles.
In this figure, the time is the execution time of the OpenCL kernel; we exclude the time required
for the tree construction and communication between CPU and GPU.
In the regime where $\Delta$ is large, the calculation time is nearly constant.
This is expected from the fact that $r_{\rm b}$ do not depend on $\Delta$.
It is evident that the setting $n_{\rm vec} = 16$ is not effective.
The setting $n_{\rm vec} = 4$ is faster in most case but 
the setting $n_{\rm vec} = 8$ outperforms it when $\Delta < 0.001$.
For a small $\Delta$, we also plot the scaling model $\propto \Delta^{-3/4}$.

\begin{figure}
\centering
\includegraphics[width=8.5cm]{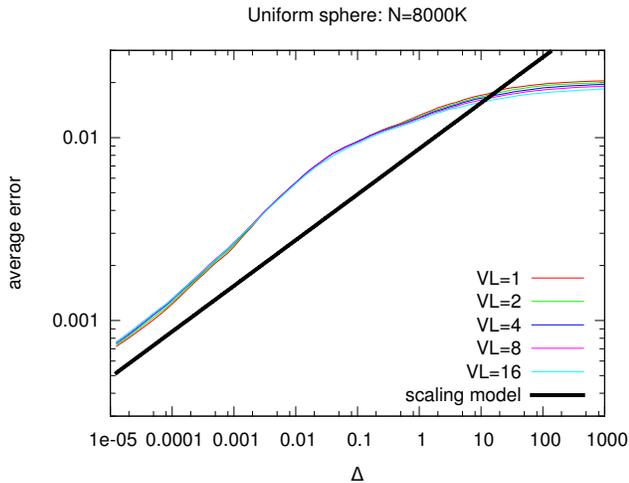}
\caption{
The relation between $\Delta$ and the average error in acceleration.
The particle distribution is the same as in Figure \ref{mac_experiment_time}.
}
\label{mac_experiment_err}
\end{figure}

Figure \ref{mac_experiment_err} shows the relation between $\Delta$ and 
the average error in acceleration.
The scaling of the average error is easily estimated in the present work
since the local error is bounded by $\Delta$ as $<a_{\rm error}> \propto \Delta^{0.25}$
for small $\Delta$.
We plot the scaling model in Figure \ref{mac_experiment_err}.
The average error do not depend on $n_{\rm vec}$ as expected.
Comparing these figures, the setting $\Delta \sim 0.1 - 0.01$ 
is optimal regarding the tradeoff between the calculation cost and the accuracy in this particular case.
With another particle distribution, over whole trend is the similar to the uniform distribution.

\subsection{Support for Multiple Devices}
In OTOO, we implemented a scheme that utilize multiple OpenCL devices.
Currently, this feature is mainly tested on 2090HAP system, which has four powerful GPU boards.
We adopt a simple domain decomposition among multiple devices
by assigning a range of particle to a OpenCL device.
All OpenCL devices use the same tree data but each device only computes
the force for particles in a given range.
In OTOO, particles are sorted by their Hilbert keys
so that we divide the $N$ particle into groups with $N/P_{\rm GPU}$ particles
where $P_{\rm GPU}$ is the number of OpenCL devices.
With the nature of the Hilbert order, the particle distribution is fairly localized.
We use OpenMP to control and synchronize $P_{\rm GPU}$ OpenCL devices.

\subsection{Summary of Optimizations}
In Table \ref{parameters}, we summarize the parameters that control
the accuracy and the performance of OTOO.
We did comprehensive benchmark tests of combinations of those parameters
and obtained the following optimal combination.
$\Delta$ controls the accuracy of the gravity force. 
Typically, we set $\Delta = 0.01$ in non-dimensional unit.
Alternatively, we can set $\Delta$ equals a fraction ($\sim 1$ \%)
of the average norm of the gravity force.
$n_{\rm crit}$ determines the maximum number of particle in a cell.
Larger $n_{\rm crit}$, we have shallower tree and shorter computing time for the tree construction
while the computing time of the tree traversal is longer.
$n_{\rm crit} = 16$ is a good setting.
$m_{\rm limit}$ alters the computing time for the tree construction and the tree traversal.
We use separate vector length for gravity and SPH force calculations
as $n_{\rm vec G}$ and $n_{\rm vec S}$.
With $n_{\rm vec G} = 4$ or $8$ and $n_{\rm vec G} = 4$, the performance is optimal on 2090HAP and 7970SB.
$P_{\rm GPU}$ is just a number of OpenCL devices, that are always GPUs in the present work.

A missing important optimization is how we distribute work load with $P_{\rm GPU} > 1$.
The simple domain decomposition we adopted works well
at least for the simulations of mergers of two white dwarf stars.
The result will be presented in Section \ref{WDmodel}.
We found no severe load imbalance in this case.
This is because we have used OTOO for at most $P_{\rm GPU} = 4$.
Practically, it will be not effective to construct a system with $P_{\rm GPU} > 4$ 
due to the limited communication bandwidth between CPU and GPU, 
e.g. 2090HAP has 80 lanes (PCI Express Gen.3).
When we will parallelize OTOO for multiple heterogenous nodes, 
we will need to consider the load balancing 
in two levels that are for GPUs and for multi nodes together.

\begin{table*}
\renewcommand{\arraystretch}{1.3}
\caption{Optimization Parameters}
\label{parameters}
\begin{center}
\begin{tabular}{c|c|c}
\hline
         & description & typical value\\
\hline
$\Delta$        & control force accuracy & 0.01 - 0.001 (non-dimensional) \\
$n_{\rm crit}$  & maximum number of particle in a cell & 8 - 32 \\
$m_{\rm limit}$ & limit mass for $B_2$                 & 0.01 - 0.05 of the total mass \\
$n_{\rm vec G}$ & vectorization factor for gravity     & 1 - 8 \\
$n_{\rm vec S}$ & vectorization factor for SPH         & 1 - 4 \\
$P_{\rm GPU}$   & number of OpenCL devices             & 1 - 4 \\
\hline
\end{tabular}
\end{center}
\end{table*}

\section{Performance Tests}
In this section, we report the performance of OTOO on various OpenCL devices
listed in Table \ref{configuration}.

\subsection{Performance of Gravity Simulations}
\begin{figure}
\centering
\includegraphics[width=8.5cm]{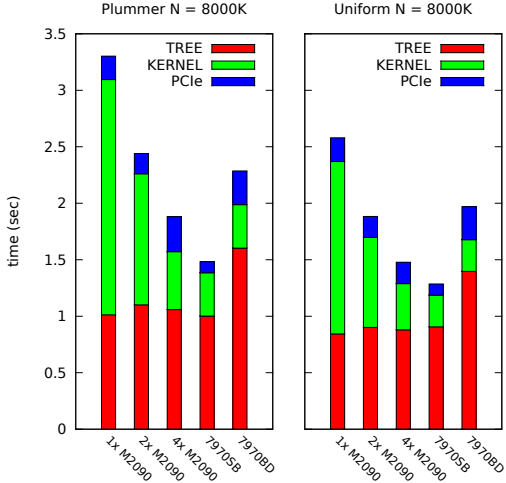}
\caption{
The left and right panels show the measured time of 
the gravity calculation on different system configurations for two particle distributions.
The labels TREE, KERNEL, PCIe show the breakdown in time spent on 
CPU (mainly tree data ), OpenCL kernel on GPU, and
the data transfer between CPU and GPU, respectively.
The configurations are presented in Table \ref{configuration}.
We put the labels ``1x'', ``2x'' and ``4x'' 
that represent $P_{\rm GPU}$ used in the 2090HAP system.
}
\label{G}
\end{figure}

First, we show that the performance of the gravity calculation.
Figure \ref{G} shows the elapsed time per a time step for $N = 8$ M models.
We used two particle distributions: (a) the Plummer model \cite{Plummer_1911} (left panel)
and (b) uniform sphere (right panel). 
The Plummer model is consistent with particle distributions typically appeared in 
astrophysics while the uniform sphere represents a distribution usually used in test of 
fast multipole method (FMM) implementations \cite{Hu_2011, Yokota_2012}.
In each panel, we present the results for 2090HAP, 7970SB and 7970BD.
Bars also shows the breakdown in time 
spent on the tree construction (red), the execution time of OpenCL kernel (green)
and the data transfer between CPU and GPU (blue).
For 2090HAP, we measured the timing with $P_{\rm GPU} = 1, 2$ and $4$. 
In this test, we set $\Delta = 0.01$, $n_{\rm vec G} = 4$, $n_{\rm crit} = 16$ and $m_{\rm limit} = 0.05$.

The time spent on the tree construction depends on the performance of CPU, i.e., 
2090HAP that has 16 cores took less than 1 sec for the tree construction
while 7970BD with 8 cores took roughly 1.5 sec.
It shows that our parallelization strategy presented in Section \ref{TC} and \ref{MM} is effective.
Scalability of the execution of the OpenCL kernel on multiple GPUs is fairly good.
For the Plummer model, 1xM2090, 2xM2090 and 4xM2090 runs took 2.08, 1.16, and 5.13 sec,respectively.
However, the time for communication is not scalable and the time for the tree construction
is nearly constant so that the total execution time shows modest scaling.
For the communication time, the effectiveness of PCI Express Gen.3 is clear;
only 7970SB fully supports Gen.3 and it took 0.1 sec for the communication 
while 7970BD with the same GPU in Gen.2 mode took 0.3 sec.
Although 2090HAP system supports Gen.3, it is currently working in Gen.2 mode with M2090 GPU boards.
To take advantage of multiple GPU devices, it will be necessary to adopt a newer GPU in HA-PACS.

\subsection{Performance Comparison of OpenCL and CUDA}
\begin{figure}
\centering
\includegraphics[width=8.5cm]{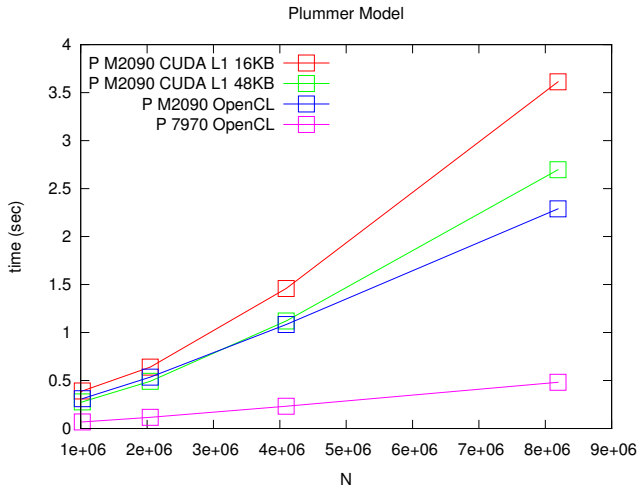}
\caption{
Comparison of the performance of CUDA and OpenCL with the Plummer particle distribution.
}
\label{CUDA_P}
\end{figure}

\begin{figure}
\centering
\includegraphics[width=8.5cm]{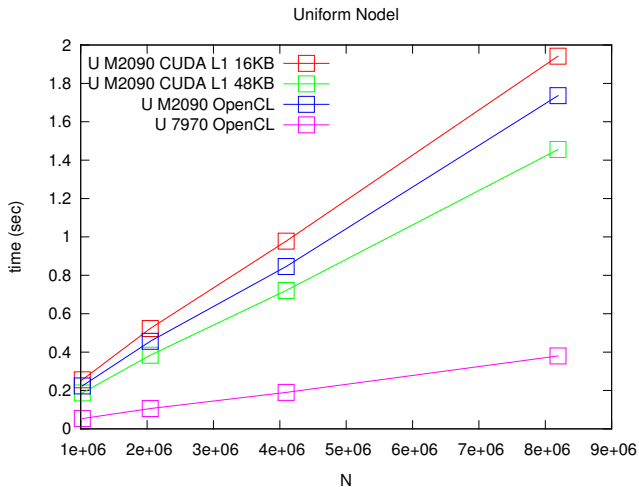}
\caption{
Comparison of the performance of CUDA and OpenCL with uniform particle distributions
}
\label{CUDA_U}
\end{figure}

On NVIDIA system, both CUDA and OpenCL is supported as a programming model.
CUDA is proprietary technology only available on GPUs by NVIDIA while
OpenCL is an open and a standard programming model.
Practically, there is little fundamental difference between two programming models.
However, a current limitation of OpenCL runtime for NVIDIA Fermi architecture
is that we can not configure the size of L1 cache memory from the default size of 16KB to 48KB.
Since our tree walk algorithm relies on the effectiveness of the cache memory system, 
it is critical for the performance of OTOO.
In addition to OTOO, we are independently implementing a CUDA enabled tree code for the Fermi architecture
so that we compare the performance of the same tree algorithm on both programming model.
To see the effectiveness of the cache, we did the following steps;
(1) we construct required tree data with OTOO for an uniform and the Plummer particle distributions.
(2) we made OTOO and the CUDA-enabled tree code loading exactly the same tree data.
(3) we measured the execution time of OTOO and the CUDA enabled code
with the same tree traversal algorithm.
We did the measurement on 2090HAP and 7970SB.
In this test, we set $\Delta = 0.01$, $n_{\rm vec G} = 4$, $n_{\rm crit} = 16$ and $m_{\rm limit} = 0.05$.
For the measurement on 2090HAP, we only used one GPU board and 
measured the performance of CUDA with the cache size of 16KB and 48KB.
The version of CUDA is 4.0.17.
Figure \ref{CUDA_P} and \ref{CUDA_U} present the performance of OTOO and the CUDA enabled tree code.

As expected, the performance of the CUDA enabled code
depends on the size of cache in both particle distributions.
With the Plummer model $N = 8$M, the runs with 16KB and 48KB took 3.61 and 2.70 sec, respectively.
OTOO that presumably uses 16KB configuration took 2.08 sec. 
With the uniform distribution $N = 8$M, the runs with 16KB and 48KB cache size,
and OTOO took 1.94, 1.46, and 1.52 sec, respectively.
Note that the particle distribution of the Plummer model is more extended $R_{\rm max} \gg 1$
while the particles in the uniform distribution are localized $R_{\rm max} = 1$
where $R_{\rm max}$ is the maximum radius of distribution.
Due to this nature of two particle distributions, 
the performance of OTOO with the Plummer model is even better than the run with 48KB.
As a comparison, the performance of OTOO on 7970SB is much better than 2090HAP.
To summarize, there is no big performance gap between OpenCL and CUDA programming models
but it will be preferable to enable a large L1 cache configuration with NVIDIA OpenCL SDK.

\subsection{Performance of Gravity+SPH runs}
Finally, we present the performance of runs with combined gravity and SPH calculations.
We did a standard test of SPH code so-called ``Cold Collapse Test'' \cite{Evrard_1988, Hernquist_1989}.
We set up an isothermal sphere with $\rho \propto r^{-2}$ distribution.
The temperature of the sphere is cold so that it collapses due to self-gravity
and eventually produces shock bound.
Here, we only report the performance of the runs till $t = 0.5$ before the shock bounce occurs
but we have checked that the later evolution obtained by OTOO is correct.
In this test, we set $\Delta = 0.01$, $n_{\rm vec G} = n_{\rm vec S} = 4$,
$n_{\rm crit} = 16$ and $m_{\rm limit} = 0.05$.

\begin{figure}
\centering
\includegraphics[width=8.5cm]{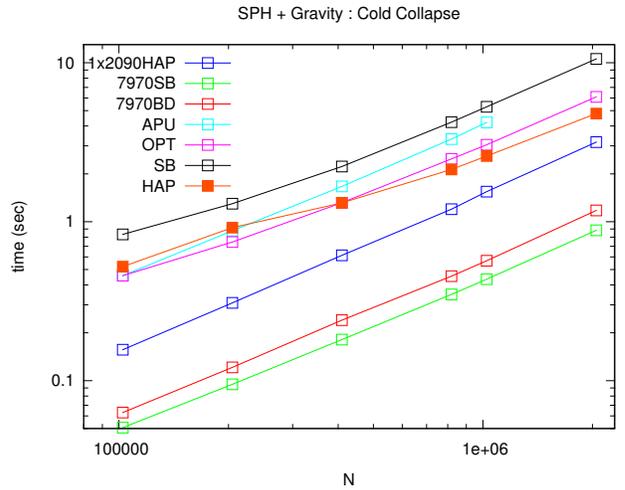}
\caption{
The execution time of gravity+SPH runs
on various systems listed in Table \ref{configuration}.
In ``1x2090HAP'', we only use one M2090 GPU board.
}
\label{SPH}
\end{figure}

Figure \ref{SPH} shows the execution time per a time step as function of $N$.
We have tested all systems listed in Table \ref{configuration}.
In all cases, the scaling on $N$ is roughly $O(N)$.
And the more peak performance, the execution time is shorter.
Systems with GPU shows relatively better performance than systems with only CPU.
We found that the heterogenous APU is competitive in comparison to systems
with high-end multi-core CPUs such as SB (6 cores), HAP(16 cores) and OPT (24 cores).
The fastest system is 7970SB, on which it took less than 1 sec with $N = 2$M run.
It is shown that OTOO is a practical tool for astrophysical modeling
on various systems that support OpenCL programming model.
We will next show a result of production runs using OTOO.

\section{Simulation of a Merger of Two White Dwarf Stars}
\label{WDmodel}
As an complex application, we have computed the evolution of a merger of two white stars with OTOO.
To model the evolution of mergers,
we implemented the following two modifications to a standard SPH method
that we have adopted in our previous code \cite{Nakasato_2003}.
First, we changed the treatment of artificial viscosity 
to minimize numerical effects according to the hybrid scheme proposed by \cite{Rosswog_2000}.
They have combined a sheer-free viscosity term \cite{Balsara_1995}
with time dependent viscosity parameters \cite{Morris_1997}.
Second, we adopted the HELMHOLTZ equation of state (EOS) \cite{Timmes_2000}.
With this EOS routine, we could calculate thermodynamic quantities as function of
temperature, density and chemical composition for mixture of 
plackian photons, an ideal gas of ions, an electron-positron gas with an arbitrary degree
of relativity and degeneracy.
Here, we assume chemical composition of white dwarf stars is 50 \% of carbon and 50 \% of oxygen.
Furthermore, we have used 2D Hermite interpolation for the HELMHOLTZ EOS.
Initial white dwarf models were constructed with a description presented in \cite{Dan_2011}.

An integration step for a merger process proceed as follows;
(1) prediction of $\vec{r}$ etc.,
(2) construct the tree structure,
(3) update the cell properties for SPH using Eq. (\ref{SPH_RAD}),
(4) tree traversal for the first stage of SPH by OpenCL,
(5) compute EOS on host,
(6) tree traversal for the second stage of SPH by OpenCL,
(7) update the cell properties for gravity using Eq. (\ref{SW_MAC}),
(8) tree traversal for gravity by OpenCL,
(9) correction of $\vec{r}$ etc.

\begin{figure}
\centering
\includegraphics[width=8cm]{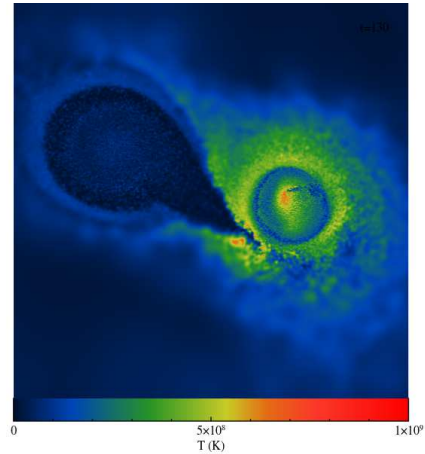}
\caption{
A snapshot of the merger of two white dwarf stars on 2090HAP with 4GPU configuration.
}
\label{WD_1}
\end{figure}

Except Steps 4, 6, and 8, all steps are executed on CPU with parallel computation as much as possible.
The steps 1 and 9 deal with the integration of particles where we adopt the leap-frog integration scheme.
In the step 4 of SPH, we compute density and divergence and rotation of velocity etc.
In the step 6 of SPH, we SPH force and time derivative of energy.
In recent results, \cite{Dan_2012} did a comprehensive survey but 
with a low resolution model $N \sim 40,000$.
\cite{Raskin_2012} used $N \sim 1,000,000$ particles and
\cite{Pakmor_2012} used $N \sim 1,800,000$ particles.
We did a production run for our work with $N = 4,096,000$ that is a largest simulation in similar modelings.
For this run, we set $\Delta = 0.01$, $n_{\rm vec G} = n_{\rm vec S} = 4$,
$n_{\rm crit} = 16$ and $m_{\rm limit} = 0.025$.
Figure \ref{WD_1} shows a snapshot of the run showing temperature distribution at the orbital plane.

This run took 246 hours on 2090HP with 4GPU configuration.
It required 251243 integration steps so that the average execution time per a step is $3.53$ sec.
We measured the performance of the same run on 7970SB that is the configuration with a latest GPU.
The average execution time is 3.20 sec with the standard g++ 4.4.3.
With Intel icpc 12.1.3, we obtained the slightly better result of 2.52 sec.
After inspecting the timing results, we found that the performance gap is due
to the difference in the efficiency of parallel regions by OpenMP.
Very roughly, the gravity calculation requires $8 \times 10^9$ particle--particle interactions
and the SPH calculation requires $3 \times 10^8$ particle--particle interactions.
The gravity and SPH interactions need 20 and 200 flops so that the effective performance
of 2090HP with 4GPU is $\sim 60$ Gflops and that of 7970SB $\sim 90$ Gflops.
For the run on 7970SB compiled with Intel icpc, 
the time spent on CPU and GPU for the 7970SB run is 1.11 and 1.41 sec, respectively.
Accordingly, we note that the effective performance of the computations on GPU is faster than
the above performance estimated from the total running time.
Notably, computationally dominant steps on CPU 
are the steps 2 and 5 that took 0.34 and 0.54 sec, respectively.
In the step 5, we compute pressure and temperature of each particles by interpolating large EOS tables.
We will need additional optimizations for those steps
for further enhancing the total performance.

Our results show that OTOO is already usable for production runs.
Even with single node, the performance of OTOO on 2090HAP and 7970SB 
is competitive in comparison with all other recent work 
that were presumably computed on a MPP with multiple nodes.
We will use OTOO to investigate the outcome of mergers of two white dwarf stars.

\section{Discussion and Summary}
In the present paper, we report the optimization strategy and detailed techniques
adopted in a simulation code OTOO for heterogenous CPU-GPU systems.
A novel part of our work is that we only use a OpenCL device for tree traversal.
In an implementation of a tree code on GPU \cite{Bedorf_2012}, 
they have proposed to construct the tree structure on GPU.
A reason to do so is that it is possible to run entire tree code, 
i.e., tree construction, update the properties of cells, and force calculation, 
on GPU without intervention by CPU.
The reported performance by \cite{Bedorf_2012} is not competitive with our performance.
And this approach is not effective when we would like to parallelize the code for a GPU cluster.
In \cite{Hu_2011}, they have reported an implementation of FMM on a heterogenous CPU-GPU cluster.
The performance with the expansion order $p = 4$ using NVIDIA C2050 took 0.37 sec for a $N = 1$ M run
while OTOO on 7970SB took 0.163 sec for $N = 1$ M uniform sphere.
In \cite{Yokota_2012}, they have proposed the hybrid FMM and tree code and reported that 
a run with $p = 5$ on NVIDIA GTX590 (dual GPU boards) 
took roughly 1 sec for a $N = 1$ M run.
Note that all those related codes have adopted the stack-based tree traversal that
is not effective in our opinion.
We believe that our algorithm is faster and more flexible since
the linked-list structure can handle any complex tree structure not limited to the octree.

We also shown that support of multiple OpenCL devices in OTOO is effective.
We did production runs of mergers of two white dwarf stars
on a recent GPU centric heterogenous cluster HA-PACS with 4GPU configurations.
We found that OTOO shows good scalability on these production runs.
As far as we know, OTOO is a first practical SPH simulation code for astrophysics
taking advantage of GPU. 
We plan to extend OTOO to support parallel runs on multiple nodes.
An advantage of a heterogenous CPU-GPU system for large scale parallel runs
is that we require less number of nodes (or MPI processes) for a given problem size
due to acceleration gained by GPU.
An important consideration for that extension is that since it takes relatively 
long time to construct tree than the gravity and SPH calculations, 
we will also need to rethink a parallel tree algorithm.
The conventional algorithm called locally essential tree \cite{Warren_1993}
that builds a global but pruned tree on each node 
will not be effective to our proposed tree method.
We will investigate the parallel tree code for a heterogenous CPU-GPU system in future work.

\bibliographystyle{IEEEtran}
\bibliography{main}

\end{document}